\documentclass{llncs}
\usepackage[T1]{fontenc}
\usepackage{graphicx}
\usepackage{cite}
\usepackage{amsmath,amssymb,amsfonts}
\usepackage{algorithmic}
\usepackage{subfig}
\usepackage{svg}
\begin{document}
\title{A Bayesian Optimization through Sequential Monte Carlo and Statistical Physics-Inspired Techniques}
%
%
\author{Anton Lebedev\inst{1} \and
Thomas Warford\inst{2} \and
M. Emre \c{S}ahin\inst{1} 
}

\authorrunning{A. Lebedev et al.}
%
\institute{The Hartree Centre, Keckwick Ln, Warrington, UK
\email{ \{anton.lebedev, emre.sahin\}@stfc.ac.uk} \and
University of Manchester,  M13 9PL, Manchester, UK
\email{thomas.warford@student.machester.ac.uk}}

\authorrunning{A. Lebedev et al.}
%

\begin{titlepage}
{\Large \textbf{Springer Copyright Notice}}

Copyright (c) 2023

This work is subject to copyright. All rights are reserved by the
Publisher, whether the whole or part of the material is concerned,
specifically the rights of translation, reprinting, reuse of illustra-
tions, recitation,broadcasting, reproduction on microfilms or in
any other physical way, and transmission or information storage
and retrieval, electronic adaptation, computer software,or by sim-
ilar or dissimilar methodology now known or hereafter developed.

\textbf{Published in}: Lecture Notes in Computer Science 14077 proceedings of the 23rd International Conference Prague, Czech Republic, July 3 - 5, 2023 Proceedings, Part V
\end{titlepage}
\maketitle              

\begin{abstract}

In this paper, we propose an approach for an application of Bayesian optimization using Sequential Monte Carlo (SMC) and concepts from the statistical physics of classical systems. Our method leverages the power of modern machine learning libraries such as NumPyro and JAX, allowing us to perform Bayesian optimization on multiple platforms, including CPUs, GPUs, TPUs, and in parallel. Our approach enables a low entry level for exploration of the methods while maintaining high performance. We present a promising direction for developing more efficient and effective techniques for a wide range of optimization problems in diverse fields.
\keywords{Stochastic Methods \and High-Performance Computing \and Bayesian Inference}
\end{abstract}

\section{Introduction}

Bayesian optimization of ever-growing models has become increasingly important in recent years and significant effort has been invested in achieving a reasonable runtime-to-solution. Unfortunately, most optimization tasks are implemented and optimized within a specific framework, resulting in a single optimized model. The proliferation of such implementations is difficult, as it requires both domain expertise and knowledge of the specific framework and programming language.

Probabilistic programming frameworks such as Stan \cite{carpenter2017stan} and (Num)Pyro \cite{bingham2019pyro}, provide such support for efficient optimisation methods such as Hamiltonian Monte Carlo (HMC) algorithm which can explore complex high-dimensional probability distributions. These frameworks are powerful tools for statistical analysis and inference.

Stan excels at handling complex hierarchical models with ease, which is often challenging in other probabilistic programming frameworks. Its user-friendly interface makes it accessible to those with little experience in Bayesian inference and statistical modelling, making it a popular choice for the accurate and efficient analysis of complex models. 
It provides domain experts with a performant tool to perform the said task with little to no programming knowledge. It achieves this by defining a "scripting language" for models and translation of these into C++ code.
It suffers, however, from a lack of inherent parallelism and a formulation of its methods in heavily-templated C++. 
NumPyro, built on JAX \cite{jax2018github}, enables efficient exploration of high-dimensional probabilistic models using different methods, while JAX itself combines the flexibility and ease-of-use of NumPy with the power and speed of hardware accelerators for efficient numerical computing. Additionally, JAX offers compatibility and portability by supporting code execution on a variety of hardware. 

Inspired by HMC and SMC descriptions in \cite{homman2014,delMoral2006}, we implemented an HMC algorithm in Python using the NumPyro and JAX frameworks and with DeepPPL \cite{baudart2021compiling} utilized to translate existing Stan models into their Python equivalents.
In this paper we present preliminary findings from our implementation of SMC for Bayesian parameter searches developed with its physical origins intact.

\section{Method Description}\label{sec:AlgorithmDescription}
Upon review of the SMC algorithm \cite{delMoral2006, homman2014} and its HMC kernel, it has become apparent that the approach resembles the cooling process of an ideal gas in a potential field with unknown minima, and that the algorithm's development would benefit from an understanding based on physical systems. To facilitate this understanding, we have undertaken a reformulation of the SMC and HMC algorithms to more accurately reflect the particle ensembles of statistical physics. 
A simple first step was the reintroduction of a temperature into the expressions
for probability, seeing as the probabilities used in these methods are the maximum-entropy
probabilities of an ensemble (collection) of particles at a fixed energy:
\begin{equation}
p_i = \frac{ e^{-E_i/(k_B T)} }{ \sum_{j=1}^N e^{-E_j/(k_B T)} }\label{eq:PhysicalProbabilities}\;.
\end{equation}
Here $k_B$ is the Boltzmann constant (carrying the dimensions of energy per degree of temperature), $T$ the temperature and $E_i = \mathcal{H}(q_i,p_i)$ is the energy of the particle $i$ at position $q_i$ with a momentum $p_i$ given a Hamiltonian:
\begin{equation}
\mathcal{H}(q,p) = \frac{p^2}{2m} + V(q)\label{eq:Hamiltonian}\;.
\end{equation}
As is common, the Hamiltonian encodes the dynamics of the system. 
Since we seek to determine the parameters that maximise the log-likelihood of the model the potential
can be defined\cite{homman2014} as:
\begin{equation}
V(q) := -\ln(P(q|X) )\label{eq:PotentialDefinition}\;,
\end{equation}
where $P(q|X)$ is the posterior probability density of the model, where $X$ is the vector containing the
observations and $q$ is the (position) vector in parameter space - the parameters of the model.

The formulation of \eqref{eq:PhysicalProbabilities}
commonly used in mathematics implies $T = \tfrac{1}{k_B}$ or an arbitrary definition of $k_B := 1$.
The former fixes a degree of freedom of the method, whilst the latter allows for a variation, but removes the 
dimensionality of the constant which, in conjunction with $T$ ensures that the argument of the exponential
remains a dimensionless number or quantity. Whilst \emph{functionally} of no consequence retaining the physical dimensions of the respective quantities allows for sanity checks of formulae during development.

The distribution of the "auxiliary momentum" described in \cite{homman2014} is naturally dependent on $T$,
with a higher temperature $T$ resulting in a broader distribution of the momenta (faster particles) and a wider area of the
initial sample space covered.

The HMC process is rather simple and described in \cite{homman2014} in sufficient detail.
In contrast, sequential Monte-Carlo is provided in a less legible form in \cite{delMoral2006}. Hence we privde here the simple
version we turn our attention to:
\begin{enumerate}
    \item Initialise the position samples to a normal distribution on $\mathbb{R}^n$ and the momentum samples to a normal distribution
    with temperature $T$.
    \item Iterate for a given number of SMC steps:
    \begin{enumerate}
        \item Propagate each particle in the ensemble using HMC.
        \item Determine the lowest energy for all particles, subtract it from every energy (renormalisation) and store it for future processing.
        \item Compute and store the average parameter value.
        \item Determine the effective size $N_{\mathit{effective}}$ of the ensemble according with \cite{delMoral2006}.
        \item If resampling is necessary select $N_{\mathit{effective}}$ particles with largest weights (lowest energies) and duplicate
        them according to their probability until the original ensemble size is reached. Then reset the momenta to a thermal distribution and the weights of the particles to $\frac{1}{N}$.
    \end{enumerate}
    \item Compute the moving average of the stored averages, using the stored energies as weights in accordance with \eqref{eq:PhysicalProbabilities}.
\end{enumerate}

Here we must note the rescaling of the weights by subtraction of the lowest energy, which is physically motivated by the freedom to choose the origin of the energy scale. Similarly, the weighted moving average in the last step results in a smoothing of the excursions of the mean after a resampling step by weighting means with large associated energies exponentially smaller (the higher the energy the more unlikely a configuration is to occur).

\section{Numerical Experiments}\label{sec:Experiments}

\subsection{Models}
\begin{figure}
\centering
\subfloat[HMC trajectories on the unconstrained $\mathbb{R}^2$ for the CT model.]{
\includegraphics[height=0.38\linewidth, width=0.47\linewidth]{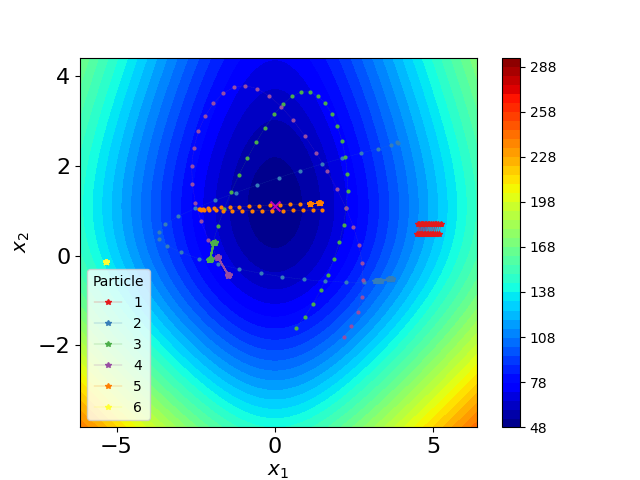}
\label{fig:HMC_Trajectories_unmapped}
}\;
\subfloat[HMC trajectories and the potential of the CT model on the unit square support of the model.]{
\includegraphics[height=0.38\linewidth, width=0.47\linewidth]{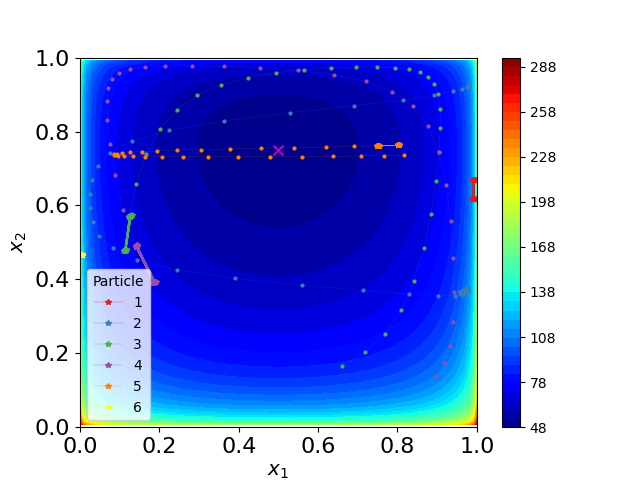}
\label{fig:HMC_Trajectories_mapped}
}
\caption{Trajectories of our HMC implementation in the potential defined by \eqref{eq:PotentialDefinition} for the CT model.
Note the dense zig-zag trajectories that result if the momentum inversion proposed in \cite{homman2014} is implemented.}
\end{figure}
To demonstrate the effectiveness of the implementation we have selected two simple models:
\begin{enumerate}
\item A sequence of $M$ independent tosses of two coins - the Coin Toss (CT) model.
\item  Item Response Model with Two-Parameter Logistic (IRT 2PL).
\end{enumerate}

\subsubsection{CT model}

The CT model assumes complete ignorance of the a-priori coin bias $p$ and its maximum a-posteriori probability estimator  can be determined formally to be:
\begin{equation}
\hat{P}_{MAP} = \frac{K}{N}\;.\label{eq:CTMAP}
\end{equation}
Here $K$ is the number of observed heads and $N = 40$ is the total number of observations.
This allows us to check the numerical approximation of \eqref{eq:PotentialDefinition} as well as its 
gradients, ensuring proper functioning of the implementation. The true parameters of the coin bias for each of the two coins are
\begin{equation}
    p_1 = \frac{1}{2}\;,\  p_2=\frac{3}{4}\;.
\end{equation}
The potential of the CT model, along with a few sample trajectories of the particles propagating therein are shown in
fig. \ref{fig:HMC_Trajectories_unmapped} prior to the constraining to the support of the model, and in fig. \ref{fig:HMC_Trajectories_mapped} after.
Here we note that we chose \emph{not} to invert the momentum in case the new phase-space point $(q,p)$ passes
the Metropolis-Hastings acceptance test, contrary to alg. 1 of \cite{homman2014}. As can be seen in the figures, such an inversion
results in a rather slow sampling of the potential in the case of a smooth potential. It is, however, beneficial for a rough potential
and hence likely better in most practical applications.

\subsubsection{IRT 2PL}
Item Response Theory (IRT) \cite{beguin2001mcmc} is a statistical model that is widely used in a variety of research fields to analyze item responses in assessments or surveys. The  two-parameter logistic (2PL) model is a specific type of IRT model that assumes each item has two parameters:  the difficulty parameter and the discrimination parameter.
The model assumes that the item responses $y_i$ for $i = 1, \ldots, I$ are Bernoulli distributed with a logit link function. The probability of responding correctly to item $i$ is given by:

\begin{equation}
P(y_i=1 | \theta, a_i, b_i) = \frac{1}{1 + \exp(-a_i(\theta - b_i))}
\end{equation}

where $\theta$ is the person's latent ability, $a_i$ is the discrimination parameter for item $i$, and $b_i$ is the difficulty parameter for item $i$. 

\subsection{Experimental Results}
Given the physical interpretation of the SMC iteration and the weighted average derived therefrom we 
expected - a-priori - a rather rapid and smooth convergence towards the true parameter values.
Given a smooth potential, as in the case of the CT model, it is furthermore expected that convergence to
the estimated parameters is faster for lower temperatures $T$.

\subsubsection{Quality of Estimation}\label{subsubsec:QoE}
\begin{figure}[tb]
\centering
\includegraphics[width=0.45\linewidth]{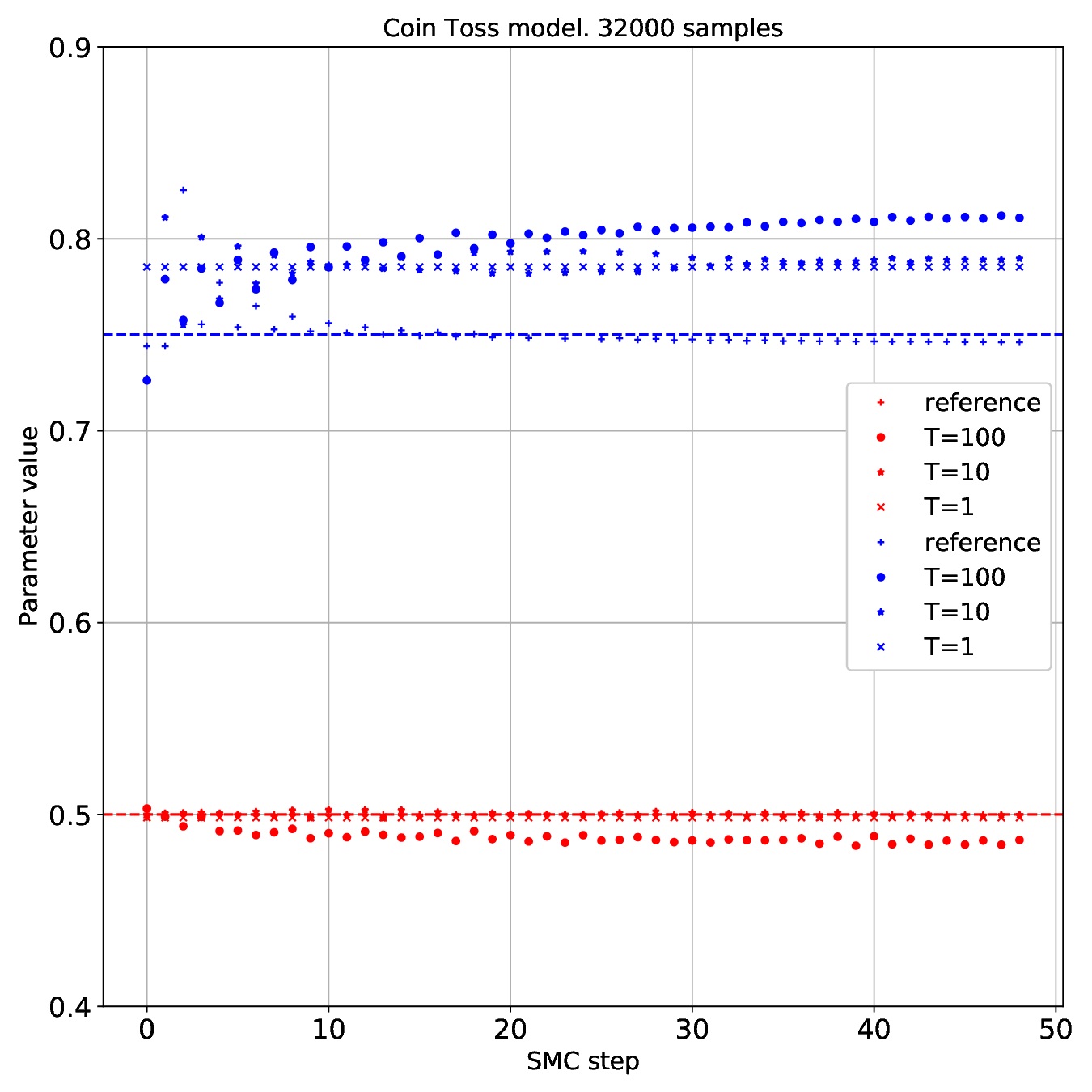}
\caption{Parameter estimates obtained with SMC with physics-motivated moving averages. The dashed lines represent the true parameter values. The crosses refer to a reference implementation not available
to the public.}
\label{fig:CT_SMC_ConvergencePlot_DifferentTemperatures}
\end{figure}
As can be seen in fig. \ref{fig:CT_SMC_ConvergencePlot_DifferentTemperatures} , all estimates of the bias of a fair coin converge within 5 iterations to
the true value $\tfrac{1}{2}$ (indicated by the dashed red line). In case of a rather high temperature (more precisely: thermal energy) of $T = 10 \frac{1}{k_B}$
the apparent limit value deviates noticeably from the true value, given the remarks above this is not unexpected and
will be remedied in a future iteration of the method.

In the case of the biased coin, with $p_2 = \frac{3}{4}$ it becomes obvious that the current development stage of the method may suffer from a freeze-in (c.f. $T=1$ case) resp. a bias in the implementation.
For a comparison reference, non-public, implementation results are marked as '+', showing a similar convergence
behaviour but without the offsets.

\subsubsection{Performance - HMC}
\begin{figure}
\centering
\subfloat[Execution time of HMC for the Coin Toss model.]{
\includegraphics[height=0.45\linewidth, width=0.47\linewidth]{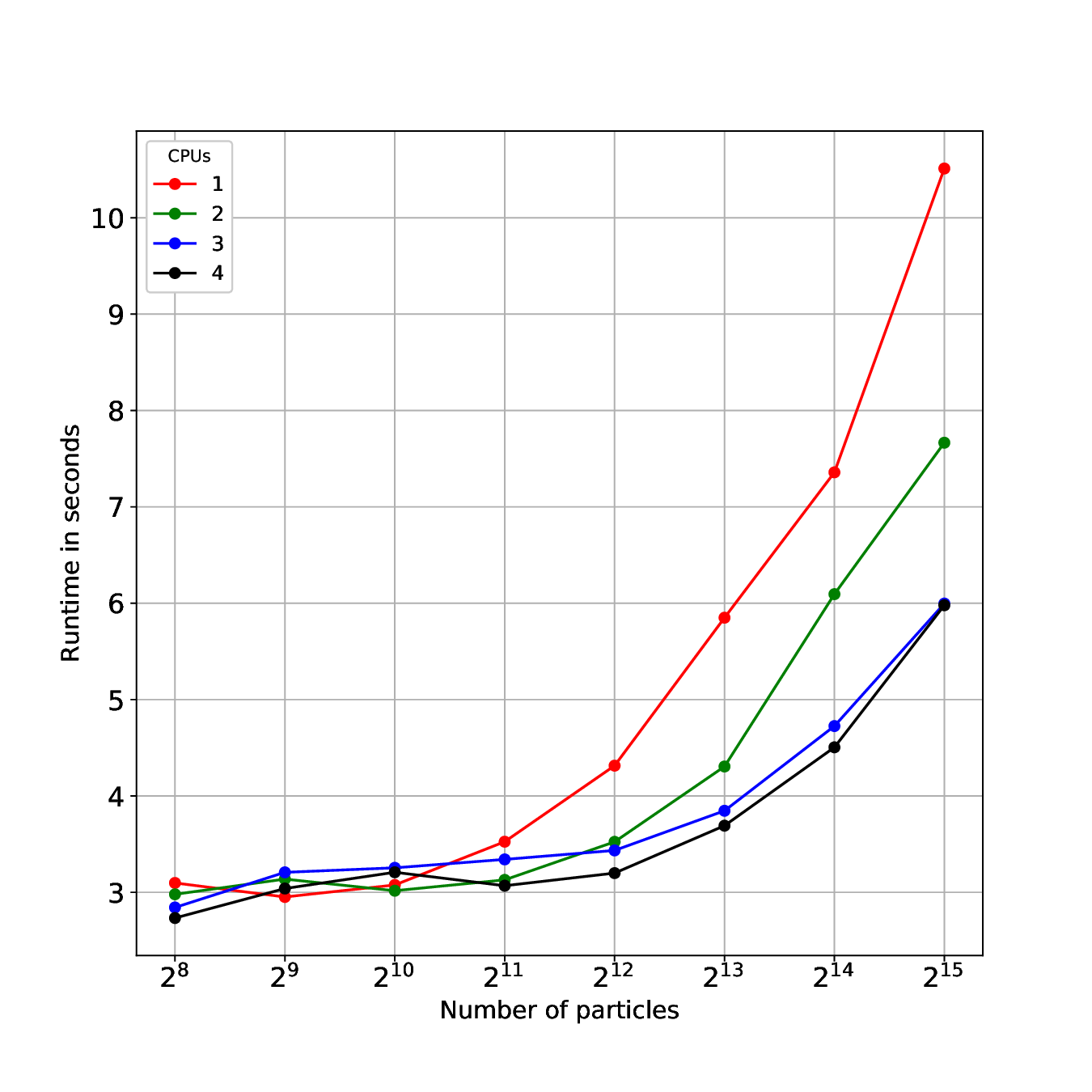}
\label{fig:CT_CPU_runtime}
}\;
\subfloat[Speed-up in comparison to $1$ CPU core for the Coin Toss model.]{
\includegraphics[height=0.45\linewidth, width=0.47\linewidth]{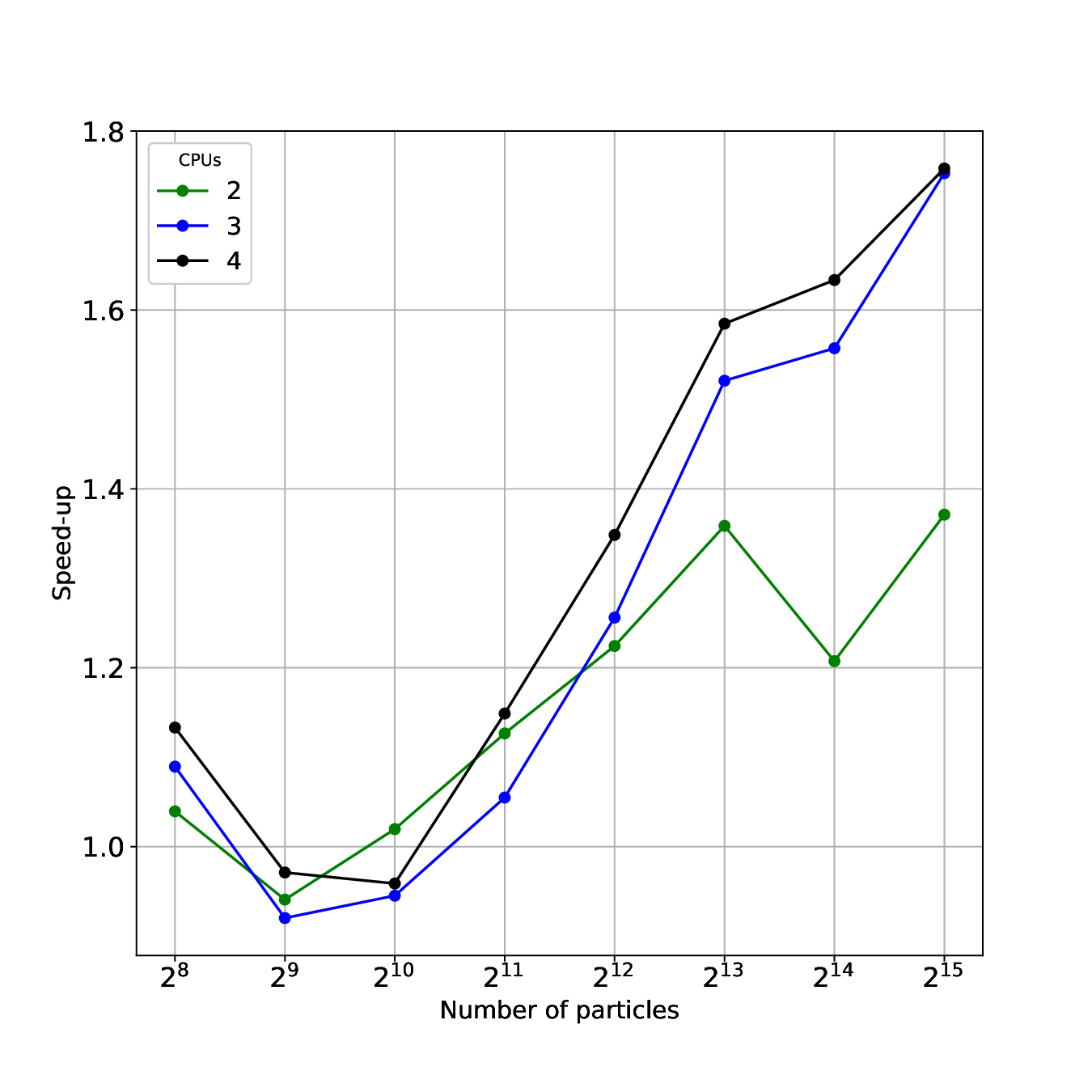}
\label{fig:CT_CPU_speedup}
}
\caption{Execution time and speed-up of our HMC implemetation when using mutliple CPU cores of a Ryzen 5 3600X to determine the coin biases of the CT model using HMC with 1000 Leap-Frog steps.}
\label{fig:CT_CPU}
\end{figure}
Although a reference implementation is available for comparison, its drawbacks are its complexity and limited execution architectures.

Our development utilises JAX and NumPyro, allowing us to run the method for variable models on a variety of architectures. Here we present the performance data obtained for the pure HMC implementation on different architectures. One can observe, in fig. \ref{fig:CT_CPU_runtime} the behaviour of the run time when running our version of HMC on multiple CPUs, parallelised via MPI. For the CT model run times for up to 2048 particles are dominated by communication and data management overhead. This is a surprisingly small number, given the simplicity of the model and the limited amount of observed data fed into it. Overall the run time growth is sub-linear up until 65538 particles for 1 CPU. The resulting speed-up of using more than one CPU is depicted in fig. \ref{fig:CT_CPU_speedup}, demonstrating a sub-linear increase even with two CPUs. This observation confirms that the model is too small to scale effectively, at least for fewer than 128k particles (c.f. the IRT model below).

\begin{figure}
\centering
\subfloat[Execution time for the IRT model.]{
\includegraphics[height=0.45\linewidth, width=0.47\linewidth]{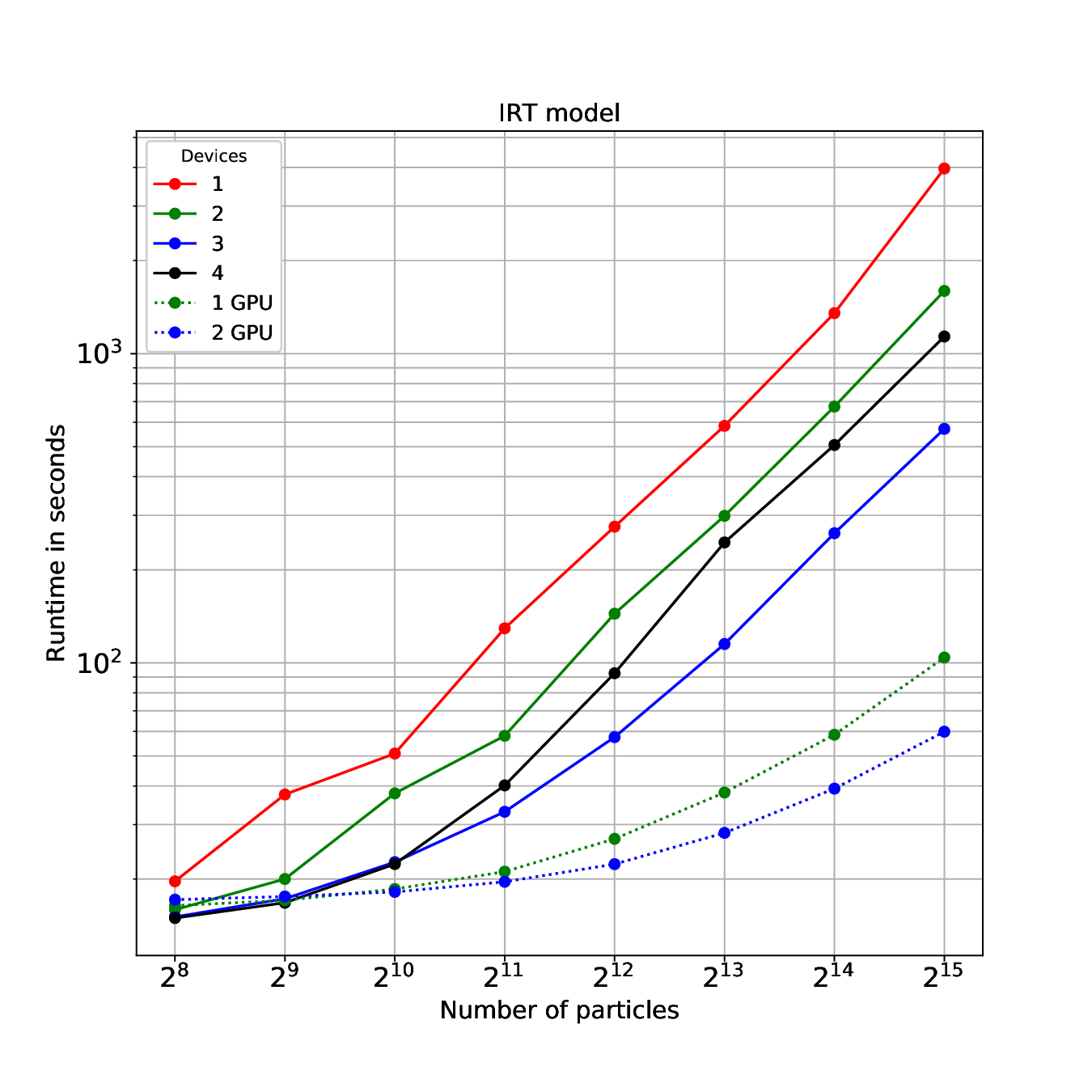}
\label{fig:IRT_CPU_GPU_runtime}
}\;
\subfloat[Speed-up in comparison to $1$ CPU core.]{
\includegraphics[height=0.45\linewidth, width=0.47\linewidth]{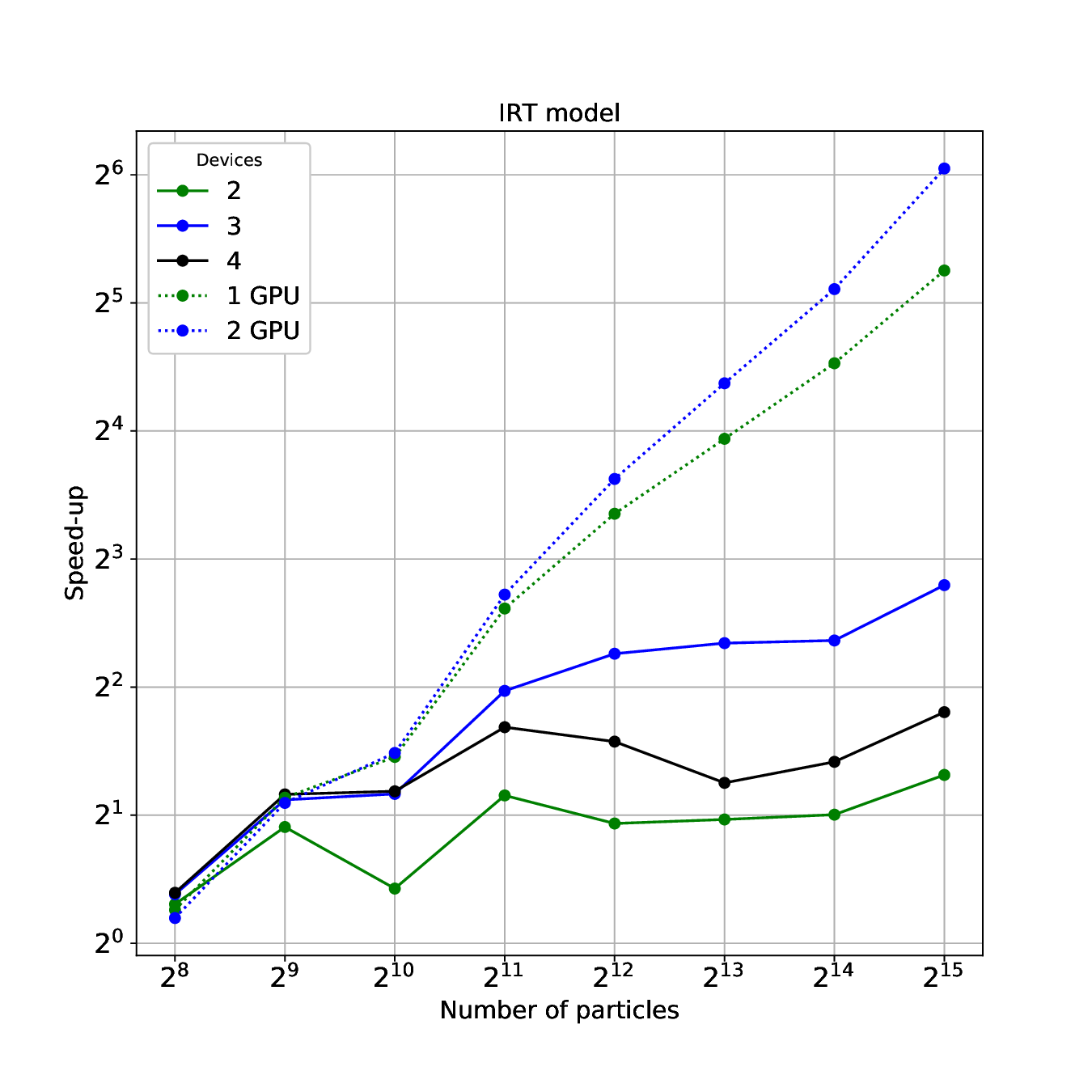}
\label{fig:IRT_CPU_GPU_speedup}
}
\caption{Execution time and speed-up of our HMC implementation when using mutliple CPU cores of a Ryzen 5 3600X as well as a RTX 3060 and GTX 1080 Ti to determine the coin biases of the IRT model for HMC with 1000 Leap-Frog steps. }
\end{figure}

In contrast to the CT model is the IRT model, whose runtimes are displayed in fig. \ref{fig:IRT_CPU_GPU_runtime}.
One can immediately see, that the run time is dominated by overhead only below 256 particles \textit{per device} (here: CPU). It is also important to note, that the run time using 4 CPU cores is larger than when only 3 are utilised. We attribute this to the apparent and unintentional spawning of multiple Python processes per MPI process. Our observations show that each MPI process spawns roughly 2 Python processes. This will be a point of future investigation. This observation explains the apparent super-linear speed-up for the CPUs displayed in fig. \ref{fig:IRT_CPU_GPU_speedup}.
One can also observe the order of magnitude reduction in the overall run time in the case of 65538 particles,
when one GPU is used. The sub-linear speed-up from 1 GPU to 2 GPUs can here be explained by the fact, that the
devices were asymmetrically bound to the system (PCIE-x16 vs. PCIE-x8) as well as that two devices of different generations were used: a GTX 1080 Ti and an RTX 3060.

The latter point shows the flexibility of JAX and NumPyro, which allowed us to run the same method on multiple CPU cores and multiple, heterogeneous, GPUs using MPI without having to modify the code!

\section{Conclusions and Future Work}

In conclusion, we have demonstrated that using the NumPyro and JAX frameworks along with intuition from classical mechanics and statistical physics, it is possible to re-create an SMC Bayesian optimisation process, whilst enriching it with an intuitive understanding of the respective steps.

The selected framework enabled us to create a simple, easy to maintain, implementation of HMC and SMC that 
can be parallelised to multiple computing devices of varying architectures on demand.
Our implementation outperforms a similar implementation in Stan by a factor of $\sim 2$ on a single CPU (core)
and scales well to multiple CPUs/GPUs, with larger gains obtained for larger models (or models with more observation
data) and more sampling particles.

In the near future, we plan to extend MPI parallelism from the core HMC stage to the entire SMC iteration, as well
as perform a thorough performance analysis and optimisation of our code, since preliminary checks indicate the existence
of, e.g., unnecessary host-device data transfers.

On the theoretical side we plan to continue the reformulation of SMC in the language of statistical physics and
expect the possibility to include maximum-entropy methods and thermalisation/annealing into the framework, utilising the
long history of MC in physics \cite{binder2019}.
Having a formulation of SMC that utilises terminology of (statistical) physics we hope to be able to
extend the approach from the classical onto the quantum domain. This holds the potential of including
existing quantum resources into Bayesian optimisation processes without requiring the user to know
the intricacies of the new architecture.

\section{Acknowledgements}
During PRACE Summer of High-Performance Computing 2022, TW was able to implement parallel HMC using the discussed formulations and included correctness checks based on physical systems and models with known closed-form solutions in the implementation.

%
%
%
 \bibliographystyle{splncs04}
 \bibliography{references}

\end{document}